\documentclass[twocolumn,showpacs,preprintnumbers,amsmath,amssymb]{revtex4}
\usepackage{epsf}
\begin{document}
  
\title{Dissipation-induced symmetry breaking in a driven optical lattice}
   
\author{R. Gommers, S. Bergamini and F. Renzoni}
   
\affiliation{Departement of Physics and Astronomy, University College London,
Gower Street, London WC1E 6BT, United Kingdom}
   
\date{\today}
   
\begin{abstract}
We analyze the atomic dynamics in an ac driven periodic optical potential 
which is symmetric in both time and space. We experimentally demonstrate 
that in the presence of dissipation the symmetry is broken, and a current of 
atoms through the optical lattice is generated as a result.
\end{abstract}
                                                                                
\maketitle

Brownian motors \cite{flashing,forced,prost94,millonas,mahato,chialvo,ratchet}
are devices which rectify the random motion of Brownian
particles, generating in this way a current. 
Quite recently a large amount of work has been devoted to the study of 
Brownian motors 
\cite{prost,marche,dykman,goychuk,blanter,luchinsky,van,flach00,flach01,super}, 
as they seem to have important applications in very different
areas. On one hand, Brownian motors may be the key for the understanding of the 
working principle of molecular motors, tiny biological engines
which transform the energy produced in chemical reactions into unidirectional
motion along periodic structures which are macroscopically flat \cite{prost}.
On the other hand, the mechanism of rectification of fluctuations identified in
the study of Brownian motors may lead to new electron pumps, and indeed the
study of solid state devices which implement Brownian motors is at present a
very active area of research \cite{linke98,linke,weiss,nori}.

The realization of a Brownian motor is a quite subtle task, and requires to 
overcome the limitations imposed by the second principle of thermodynamics 
and to break the symmetries which inhibit directed motion. Indeed, as the 
second principle of thermodynamics does not allow the appearance of a current 
in a system at thermal equilibrium, Brownian motors are realized by driving 
Brownian particles out of equilibrium, as identified in the early proposals
for flashing \cite{flashing} and rocking \cite{forced,prost94} ratchets. 
In order then to obtain directed motion in the system out of equilibrium, 
relevant symmetries have to be broken.

Theoretical work \cite{prost94,flach00} clearly identified the symmetries 
which in the Hamiltonian limit, i.e. in the absence of dissipation, inhibit 
directed motion. However, in the presence of dissipation the scenario may 
change \cite{flach00,flach01,super} and 
it was theoretically shown that the symmetry properties of the system 
are modified by the presence of dissipation and a current can be generated
even when the symmetries of the Hamiltonian would prevent current generation 
in the Hamiltonian limit \cite{flach00,flach01}.

In this work we demonstrate experimentally the phenomenon of 
dissipation-induced symmetry breaking using cold atoms in an ac driven 
periodic optical potential \cite{robi} which is symmetric in both time 
and space. We show that in the presence of dissipation the symmetry is
broken, and a current of atoms through the optical lattice is generated 
as a result.

Before describing our experimental observations, it is essential to introduce
the relevant symmetries which control the current generation in the Hamiltonian
limit \cite{prost94,flach00,flach01}. We consider a particle in a spatially
symmetric periodic potential, periodically rocked by a zero-mean force $F(t)$
of period $T=2\pi/\omega$. In this case there are two relevant symmetries which
need to be examined to determine whether current generation is possible.
Following the notations of Ref.~\cite{flach01} we say that $F(t)$ possesses 
$\hat{F}_s$ symmetry if $F(t)$ is symmetric, after some appropriate shift: 
$F(t+\tau)=F(-t+\tau)$. Moreover, if $F(t)$ satisfies
$F(t)=-F(t+T/2)$, we say that $F$ possesses $F_{sh}$ symmetry. The symmetry 
$\hat{F}_{sh}$ of the driving implies that the system is invariant
under the shift-transformation $\hat{S}_a: (x,p,t)\to (-x,-p,t+T/2)$, while
the symmetry $\hat{F}_s$ leads to invariance under the time-reversal 
transformation $\hat{S}_b: (x,p,t)\to (x,-p,-t)$. The invariance of the system
under any of these two transformations forbids the appearance of a directed 
current \cite{flach00,flach01}. To elaborate further, we consider the specific
form for $F(t)$:
\begin{equation}
F(t)=F_0 [ A\cos\omega t +B \cos (2\omega t+\phi)]~.
\label{force}
\end{equation}
The presence of both harmonics ($F_0,A,B\neq 0$) breaks the shift symmetry
$F_{sh}$, independently of the value of the relative phase $\phi$. On the 
other hand, whether or not the $F_s$ symmetry is broken depends on the 
value of the phase $\phi$: for $\phi=n\pi$, with $n$ integer, the symmetry
$F_s$ is preserved, while for $\phi\neq n\pi$ it is broken. Therefore for 
$\phi=n\pi$ current generation is forbidden in the Hamiltonian limit, while 
for $\phi\neq n\pi$ is allowed. Perturbative calculations \cite{flach00} show 
that the average current of particles is, in leading order, proportional to
$\sin\phi$, in agreement with the symmetry considerations discussed above.
The limit in which the effect of dissipation on the symmetries is negligible 
was already experimentally examined in Ref.~\cite{michele} where the dependence
$I\sim \sin\phi$ was demonstrated.

Consider now the case of weak, nonzero dissipation. The presence of dissipation
breaks the invariance under time-reversal transformation $\hat{S}_b$ even if 
the driving is $\hat{F}_s$-symmetric. Therefore for a biharmonic force of 
the form of Eq.~(\ref{force}) both time-reversal and shift symmetries are
violated in the presence of dissipation. A current can be generated as a result. 
An interesting issue is how the dependence of the current on the phase $\phi$ 
is modified by the presence of dissipation. Calculations done by solving
the kinetic Boltzmann equation for an ensemble of interacting particles
\cite{flach01} showed that in the presence of weak dissipation the average 
current of particles $I$ still shows an approximate sinusoidal dependence 
on the phase $\phi$, but a phase lag $\phi_0$ appears as a result of 
dissipation:  
\begin{equation}
I\sim\sin(\phi-\phi_0)~. 
\label{phase_lag}
\end{equation}
The phase lag $\phi_0$ vanishes in the Hamiltonian limit and is an increasing 
function of the relaxation rate \cite{flach01}. This dependence is consistant 
with the previous observation that weak dissipation breaks the time-reversal
symmetry of the system and leads to the generation of current also for 
$\phi=n\pi$. In other words, the shift $\phi_0$ is a signature of the 
phenomenon of dissipation-induced symmetry breaking, and in our experiment
$\phi_0$ will precisely be the quantity examined to detect the phenomenon.

In our experiment we demonstrate the phenomenon of dissipation-induced 
symmetry breaking by
using cold atoms in an ac driven near-resonant optical lattice \cite{robi}. 
This is the same system used previously to demonstrate the rectification of 
fluctuations \cite{phil} and to investigate the phenomenon of resonant 
activation \cite{ralf}. These investigations clearly showed that near-resonant
optical lattices represent an ideal model system to investigate phenomena of 
statistical physics. In fact, in near-resonant optical lattices the laser 
fields create both a periodic potential for the atoms and introduce 
dissipation. More precisely, the interference of the laser fields creates one 
periodic potential for each ground state of the atoms. The laser fields also
introduce stochastic transitions (optical pumping processes) between different
ground states. This leads to damping, an effect named Sisyphus cooling, and a
fluctuating force. As a result of the fluctuations, the atoms undergo a random
walk through the periodic potential, and indeed normal diffusion was observed 
for atoms in dissipative optical lattices \cite{regis}.

The experimental set-up is the same as the one used in our previous work 
\cite{phil,ralf}. Cesium atoms are cooled and trapped in a 3D optical lattice
created by four laser beams arranged in the so-called umbrella-like
configuration. One beam (beam 1) propagates in the $z$-direction; the three 
other beams (beams 2--4) propagate in the opposite direction, arranged along
the edges of a triangular pyramid having the $z$ direction as axis. For further
detail on the lattice beams arrangement, as lattice beams' angles and
polarizations, we refer to Ref.~\cite{phil}. A zero-mean oscillating force of
the form (\ref{force}) can be applied by phase modulating one of the lattice 
beams. More precisley \cite{phil} a phase modulation of beam 1 of the form
\begin{equation}
\alpha (t) = \alpha_0 [A\cos\omega t + \frac{B}{4}\cos (2\omega t-\phi)]
\end{equation}
will result, in the accelerated frame in which the optical lattice is 
stationary, in an homogeneous force in the $z$ direction of the form of 
Eq.~(\ref{force}) with 
$F_0=m\omega^2\alpha_0/2k$, where $m$ is the atomic mass and $k$ the laser
field wavevector.

Before describing the experimental results, it is necessary to analyze 
theoretically our system to determine whether the description of the 
dissipation-induced symmetry breaking in terms of a phase lag $\phi_0$
derived for an ensemble of particles in the presence of collisions 
\cite{flach01} 
applies also to the present case of non-interacting atoms, with the 
dissipation associated to the scattering of photons. For the sake of 
simplicity, and to make the analysis more transparent, we consider the 
simplest atom-light configuration in which Sisyphus cooling takes 
place: a $J_g=1/2\to J_e=3/2$ atomic transition, and a 1D light configuration
consisting of two linearly polarized laser fields, counterpropagating and 
with orthogonal polarizations -- the so called lin$\perp$lin configuration 
\cite{robi}.
This atom-light configuration results in two optical potentials $U_{\pm}$
for the atoms, one for each ground state $|\pm\rangle$, in phase opposition:
$U_{\pm}=U_0[-2\pm \cos 2kz ]/2$, where $z$ is the atomic position along
the axis $Oz$ of light propagation and $U_0$ is the depth of the optical
potential. The damping arises from stochastic transitions between the two
optical potentials $U_{\pm}$. These stochastic transitions correspond to the 
absorption and subsequent spontaneous emission of photons. Quantitatively, 
the departure rates $\gamma_{\pm\to \mp}(z)$ from the $|\pm\rangle$
states can be written in terms of the photon scattering rate
$\Gamma'$ as $\gamma_{\pm\to \mp}= \Gamma' (1\pm \cos 2kz) /9$ \cite{robi}.
The level of dissipation can therefore be quantitatively characterized by
the photon scattering rate $\Gamma'$, which can be controlled experimentally
by varying the lattice fields parameters.

%%%% here fig. 1
%%%%%%%%%%%%%%%%%%%%%%%%%%%%%%%%%%%%%%%%%%%%%%%%%%%%%%%%%%%%%%%%%%%%%%%%%%%
\begin{figure}[ht]
\begin{center}
\mbox{\epsfxsize 3.in \epsfbox{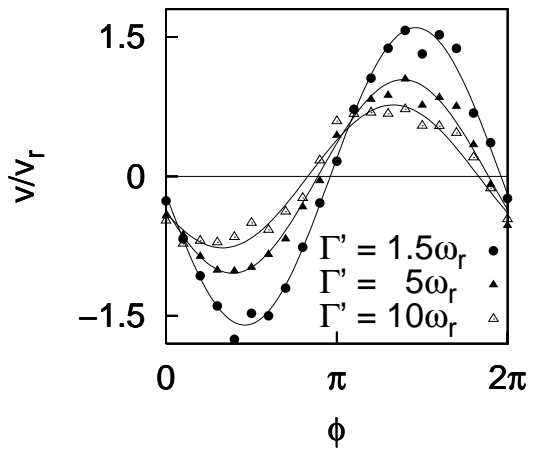}}
\end{center}
\caption{
Results of semiclassical Monte Carlo simulations for a sample of $n=10^4$
atoms in an ac driven 1D lin$\perp$lin optical lattice. The average atomic
velocity, in units of the atomic recoil velocity $v_r=\hbar k/m$, is shown
as a function of the phase difference $\phi$ between the two harmonics of the
driving force, see Eq.~(\protect\ref{force}). Different data sets correspond
to different scattering rates $\Gamma'$, expressed in units of the recoil
angular frequency $\omega_r$. The lines are the best fits of the
data with the function $v/v_r=A\sin (\phi-\phi_0)$. The calculations were
done for a lattice with depth $U_0=100 E_r$. The parameters of the driving are:
$A=1.5$, $B=6$, $\alpha_0=0.2\cdot\pi$, $\omega=0.75\omega_v$, where $\omega_v$
is the vibrational frequency of the atoms at the bottom of the wells.
}
\label{fig1}
\end{figure}

\begin{figure}[ht]
\begin{center}
\hspace*{-2cm}\mbox{\epsfxsize 3.5in \epsfbox{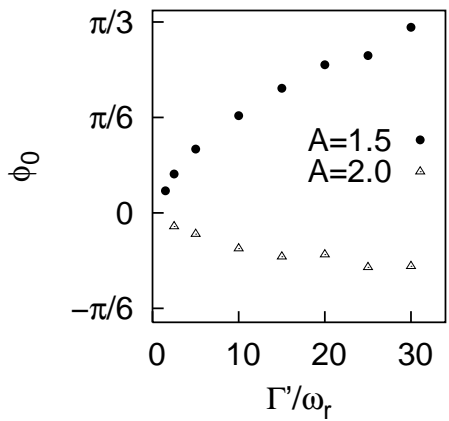}}
\end{center}
\caption{
Numerically calculated phase shift $\phi_0$ as a function of the scattering
rate $\Gamma'$. The phase shift is determined by fitting data as those
in Fig.~\protect\ref{fig1} with the function $v/v_r=A\sin(\phi-\phi_0)$.
Different data sets correspond to different driving strength, i.e. to different
values of A, with the ratio of the strengths of the two harmonics kept fixed:
$B=4 A$. All the other parameters are the same as for Fig.~\protect\ref{fig1}.
}
\label{fig2}
\end{figure}

We studied the atomic dynamics in the presence of bi-harmonic driving by
semi-classical Monte-Carlo simulations \cite{robi}. For a given optical
potential depth $U_0$, we calculated the average atomic velocity $v$ as a
function of the phase difference $\phi$ between driving fields, for different
values of the scattering rate. The results of our calculation are shown in
Fig.~\ref{fig1}. The atomic current shows a dependence of the type of
Eq.~\ref{phase_lag}, with the phase lag $\phi_0$ vanishing in the
Hamiltonian limit ($\Gamma'\to 0$) and increasing for increasing scattering
rate. Values for $\phi_0$ as a function of the scattering rate, determined
by fitting  data as those of Fig.~\ref{fig1} with $A\sin(\phi-\phi_0)$,
are reported in Fig.~\ref{fig2} for two different values of the driving force
amplitude. It can be seen that although the magnitude and sign of the phase
lag $\phi_0$ are different for the two different driving strengths considered,
the general behavior described above is observed in both cases. For
completeness, we mention here that also the sign and magnitude of the amplitude
$A$ varies depending on the driving strength. We notice that the dependence
of the sign of $\phi_0$ and $A$ on the driving strength is consistant with 
previous observations of current reversals as a function of the driving 
amplitude at a given relative phase $\phi$ between the two driving fields
\cite{phil}.

These results for the phase lag $\phi_0$ are in agreement with general symmetry
considerations \cite{flach00} and also extend the validity of the
theoretical results obtained for an ensemble of interacting particles 
\cite{flach01} to the present system of non-interacting atoms, with the 
dissipation associated to the scattering of photons.

%%%%%%%%%% here fig. 2
%%%%%%%%%%%%%%%%%%%%%%%%%%%%%%%%%%%%%%%%%%%%%%%%%%%%%%%%%%%%%%%%%%%%%%%%%%%

The experimental procedure followed closely the approach of our numerical 
simulations. Pre-cooled cesium atoms are loaded in a near-resonant optical 
lattice. The phase modulation $\alpha(t)$ is then slowly turned on. By direct
imaging the atomic cloud with a charge-coupled device camera, we then 
derived the average atomic velocity. By repeating the experiment for different
values of the phase difference $\phi$, we determined the average atomic 
velocity as a function of the phase $\phi$.

Different sets of measurements were taken for different values of the
scattering rate $\Gamma'$ at a constant depth of the optical potential. 
This was done by varying simultanously the intensity $I_L$ and detuning
$\Delta$ of the lattice beams, so to keep the potential depth 
$U_0\propto I_L/\Delta$ constant while varying the scattering rate 
$\Gamma'\propto I_L/\Delta^2$. We notice that as $I_L$ and $\Delta$ can be 
varied only within a finite range, we cannot suppress completely dissipation,
i.e. obtain $\Gamma'=0$. However, as we will see, for the driving strength
considered in the experiment, the smallest accessible scattering rate results
in a phase shift which is zero within the experimental error, i.e. this choice
of parameters well approximates the dissipationless case. By then increasing
$\Gamma'$ it is possible to investigate the effects of dissipation.

%%%%%%%%%%%%%%%%%%%%%%%%%%%%%%%%%%%%%%%%%%%%%%%%%%%%%%%%%%%%%%%%%%%%%%%%%%%
\begin{figure}[ht]
\begin{center}
\mbox{\epsfxsize 3.in \epsfbox{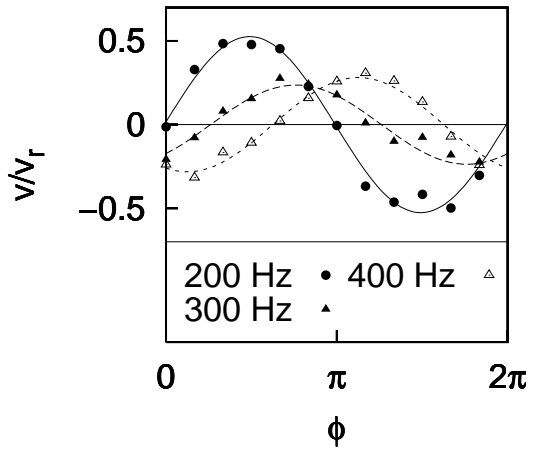}}
\end{center}
\caption{
Experimental results for the average atomic velocity, in units of the
recoil velocity, as a function of the phase $\phi$. The lines are the
best fit of the data with the function $v/v_r=A\sin(\phi-\phi_0)$. The
optical potential is the same for all measurements, and corresponds to
a vibrational frequency $\omega_v=2\pi\cdot 170$ kHz. Different data
sets correspond to different lattice detuning $\Delta$, i.e. to different
scattering rates as the optical potential is kept constant. The data
are labeled by the quantity $\Gamma_s=[\omega_v/(2\pi)]^2/\Delta$ which is
proportional to the scattering rate. The parameters of the driving
are $\omega=2\pi\cdot 100$ kHz, $A=1$, $B=4$, $\alpha_0=27.2 {\rm rad}$.
The errors on $v/v_r$ are of the order of 0.05.}
\label{fig3}
\end{figure}

\begin{figure}[ht]
\begin{center}
\hspace*{-2.5cm}\mbox{\epsfxsize 4.in \epsfbox{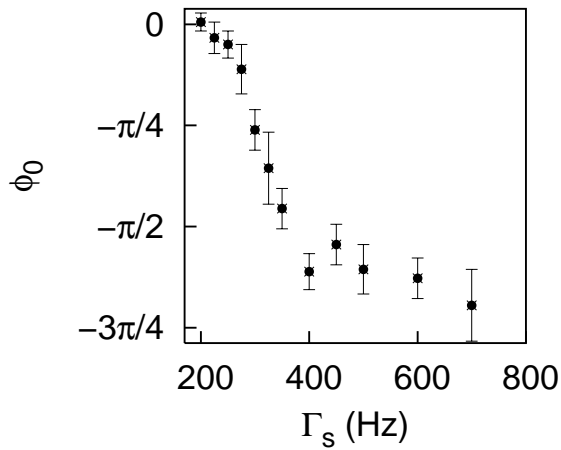}}
\end{center}
\caption{Experimental results for the phase shift $\phi_0$ as a function
of $\Gamma_s=[\omega_v/(2\pi)]^2/\Delta$, which is proportional to the
scattering rate. All the other parameters are kept constant, and are the
same as for Fig. \protect\ref{fig3}.
}
\label{fig4}
\end{figure}

The results of our measurements, reported in Fig.~\ref{fig3}, demonstrate 
clearly the phenomenon of dissipation-induced symmetry-breaking. In agreement
with our numerical calculations and with previous theoretical work 
\cite{flach00,flach01}, the measured current of atoms is well approximated
by $A\sin(\phi-\phi_0)$. Therefore, by fitting data as those reported in 
Fig.~\ref{fig3} with the function $v/v_r=A\sin(\phi-\phi_0)$ we were able 
to determine the phase shift $\phi_0$ as a function of $\Gamma'$, as 
reported in Fig.~\ref{fig4}.  The measured phase shift $\phi_0$ is zero,
within the experimental error, for the smallest scattering rate examined in 
the experiment.  In this case, no current is generated for $\phi=n\pi$, with
$n$ integer, as for this value of the phase the system is invariant under 
time-reversal transformation.  The magnitude of the phase shift $\phi_0$ 
increases at increasing scattering rate, and differs significantly from zero.
The nonzero phase shift corresponds to current generation for $\phi=n\pi$, 
i.e. when the system Hamiltonian is invariant under the time-reversal 
transformation.  This clearly demonstrates the breaking of the system 
symmetry by dissipation.

%%%%%%%%% fig. 4
%%%%%%%%%%%%%%%%%%%%%%%%%%%%%%%%%%%%%%%%%%%%%%%%%%%%%%%%%%%%%%%%%%%%%%%%%%%

In conclusion, we demonstrated experimentally the phenomenon of 
dissipation-induced symmetry breaking with cold atoms in an optical lattice.
We analyzed the atomic dynamics in an ac driven periodic optical potential
which is symmetric in both time and space. These symmetries forbid the 
generation of a current. We showed that in the
presence of dissipation the symmetry is broken, and a current of atoms
through the optical lattice is generated as a result. Our results also 
show the generality of the phenomenon, in particular extending
the results \cite{flach01} previously obtained for an ensemble of 
interacting particles in the specific framework of the kinetic Boltzmann 
equation to a system in which the dissipation mechanism is of completely
different nature. The present work is of relevance for the research on 
control of transport by time-dependent fields in a variety of system, ranging
from optical tweezer set-ups \cite{grier} to quantum dots and wires
\cite{kohler}.

We thank EPSRC, UK and the Royal Society for financial support.

\end{document}